\documentclass[letter,twocolumn]{jpsj3}
\usepackage{txfonts}
\usepackage{color}
\usepackage[resetlabels,labeled]{multibib}

\title{Superconducting order parameter in UTe$_2$ determined by Knight shift measurement}
\author{Hiroki Fujibayashi$^1$\thanks{fujibayashi.hiroki.57z@st.kyoto-u.ac.jp}, Genki Nakamine$^{1}$, Katsuki Kinjo$^{1}$, Shunsaku Kitagawa$^{1}$, Kenji Ishida$^1$\thanks{kishida@scphys.kyoto-u.ac.jp}, \\Yo Tokunaga$^{2}$, Hironori Sakai$^{2}$, Shinsaku Kambe$^{2}$, Ai Nakamura$^{3}$, Yusei Shimizu$^{3}$, \\Yoshiya Homma$^{3}$, Dexin Li$^{3}$, Fuminori Honda$^{3}$, and Dai Aoki$^{3,4}$}
\inst{
$^1$Department of Physics, Graduate School of Science, Kyoto University, Kyoto 606-8502, Japan \\
$^2$Advanced Science Research Center, Japan Atomic Energy Agency, Tokai, Ibaraki 319-1195, Japan \\
$^3$Institute for Materials Research, Tohoku University, Oarai, Ibaraki 311-1313, Japan\\
$^4$Universit\'e Grenoble Alpes, CEA, IRIG, PHELIQS, F-38000 Grenoble, France} 
\date{\today}
\abst{
This study investigates the spin susceptibility in U-based superconductor UTe$_2$ in the superconducting (SC) state by using Knight shift measurements for a magnetic field $H$ along the $a$ axis, which is the magnetic easy axis of UTe$_2$.
Although a tiny anomaly ascribed to the SC diamagnetic effect was observed just below the SC transition temperature $T_{\rm c}$, the $a$-axis Knight shift in the SC state shows no significant decrease, following the extrapolation from the normal-state temperature dependence.
This indicates that the spin susceptibility is nearly unchanged below $T_{\rm c}$.
Considering the previous Knight shift results for $H \parallel b$ and $H \parallel c$, the dominant SC state is determined to be $B_{\rm 3u}$ in the spin-triplet pairing, which is consistent with the spin anisotropy in the normal state.
The present result shows that UTe$_2$ is a spin-triplet superconductor with spin degrees of freedom.}

\begin{document}
\maketitle
UTe$_2$ is a superconductor with a superconducting (SC) transition temperature ($T_{\rm c}$) of $1.6 \sim 2.0$~K\cite{RanScience2019,AokiJPSJ2019,RosaCondMat2021}. 
Although UTe$_2$ does not exhibit ferromagnetism, it is considered as an end member of ferromagnetic (FM) superconductors owing to their similar physical characteristics, such as Ising anisotropy in the magnetic susceptibility\cite{RanScience2019,AokiJPSJ2019}, the phenomenon of SC upper critical field ($H_{\rm{c2}}$) in all the crystal axes exceeding the limit of Pauli-depairing effect\cite{RanScience2019,AokiJPSJ2019},and superconductivity boosted by a magnetic field ($H$) above 20~T for $H \parallel b$ (magnetic hard axis)\cite{RanNatPhy2019,KnebelJPSJ2019}, etc.
Thus, UTe$_2$ is anticipated to be a spin-triplet superconductor, similar to FM superconductors.
The experimental results of multiple SC phases\cite{BraithwaiteCommPhy2019,ThomasSciAdv2020}, spontaneous broken time-reversal symmetry\cite{HayesScience2021}, and the chiral Majorana edge and surface state\cite{JiaoNature2020} suggest spin-triplet superconductivity with spin and/or orbital degrees of freedom.

\begin{figure}[tbp]
\begin{center}
\includegraphics[width=8.5cm]{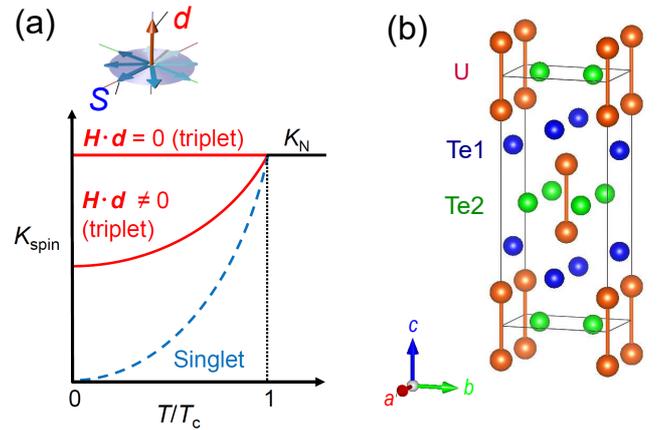}
\end{center}
\caption{(Color online) (a) Image of the $\mbox{\boldmath$d$}$-vector. Here $\mbox{\boldmath$S$}$ denotes the spin component of the spin-triplet pairing. 
Knight-shift behavior in the SC state. 
The behavior is different in a spin-triplet superconductor, depending on the relationship between $\mbox{\boldmath$d$}$ and the magnetic-field vector $\mbox{\boldmath$H$}$. 
In contrast, the Knight shift decreases in all directions in a spin-singlet superconductor. 
(b) The crystal structure of UTe$_2$.       }
\label{f1}
\end{figure}

We measured the $^{125}$Te-NMR  Knight shifts of UTe$_2$ to investigate the spin susceptibility in the SC state\cite{NakamineJPSJ2019, NakaminePRB2021, NakamineJPSJ2021} because the measurement of the Knight shift probing the static field  at the nuclear site is one of the most reliable methods for measuring spin susceptibility in the SC state.
In a spin-triplet superconductor with spin degrees of freedom, the spin state in the SC state is characterized by the $\mbox{\boldmath$d$}$ vector perpendicular to the spin component of the spin-triplet pairing \cite{LeggettRMP1975}. 
If the $\mbox{\boldmath$d$}$ vector is fixed to one of the crystalline axes by any mechanism, the spin component of the Knight shift decreases in the SC state when $\mbox{\boldmath$d$}$-vector has a finite component of the magnetic field direction  ($\mbox{\boldmath$H$} \cdot \mbox{\boldmath$d$} \neq 0$); however, it remains unchanged when $\mbox{\boldmath$H$} \perp \mbox{\boldmath$d$}$ ($\mbox{\boldmath$H$} \cdot \mbox{\boldmath$d$} = 0$).
This relation is illustrated in Fig.~\ref{f1}(a). 
Thus, the direction of the $\mbox{\boldmath$d$}$ vector can be derived from the measurements of the Knight shift along each crystalline axis.
In addition, based on the orthorhombic crystal structure of UTe$_2$, the SC symmetry is classified as a $D_{\rm 2h}$ point group, and the irreducible representation for spin-triplet pairing is listed in Table \ref{t1} \cite{IshizukaPRL2019}.  

\begin{table}[htb]
\begin{center}
\caption{\label{t1}Classification of the odd-parity SC order parameters for point groups with $D_{\rm 2h}$. The irreducible representation (IR) and its basis functions are listed. To clarify the dominant SC spin component, a spin component perpendicular to the $\mbox{\boldmath$d$}$ vectors is also shown.}
\vspace{5mm}
  \begin{tabular}{cccc}\hline \hline 
 \multicolumn{1}{c}{$D_{\rm 2h}$ (zero field)} \\ \hline 
IR & Basis functions & & SC spin comp.\\ \hline
$A_{\rm u}$ & $k_a \hat{a}$, $k_b \hat{b}$, $k_c \hat{c}$ & & \\
$B_{\rm 1u}$ & $k_b \hat{a}$, $k_a \hat{b}$ & & $c$ \\
$B_{\rm 2u}$ & $k_a \hat{c}$, $k_c \hat{a}$ & & $b$ \\
$B_{\rm 3u}$ & $k_c \hat{b}$, $k_b \hat{c}$ & & $a$ \\ \hline \hline
\\ \\
  \end{tabular}
  \end{center}
\end{table}

In previous studies\cite{NakamineJPSJ2019, NakaminePRB2021}, we have reported a slight decrease in the Knight shift along the $b$ and $c$ axes ($K_b$ and $K_c$, respectively) at a low magnetic field of $\mu_0 H$ = 1 T in the SC state.
The decreases in $K_b$ and $K_c$ are smaller than those expected in the spin-singlet pairing. 
This indicates that the components of $\hat{b}$ and $\hat{c}$ in the $\mbox{\boldmath$d$}$ vector are finite.  
In addition, we found that the $H$ response of the $\mbox{\boldmath$d$}$ vector is anisotropic for the $b$ and $c$ axes\cite{NakamineJPSJ2021}: the $\hat{c}$ component is quickly suppressed with increasing field for $H \parallel c$, but the $\hat{b}$ component, which is constant at low $H$, begins to decease above 7~T and almost zero at 12.5~T for $H \parallel b$.
As 12.5 T is not far from the inflection of $H_{\rm c2}$ ($\mu_0 H_{\rm c2} \sim 15$ T), we suggested that the direction of the $\mbox{\boldmath$d$}$ vector would be different between the low-$H$ and $H$-boosted high-$H$ SC phases for $H \parallel b$\cite{NakaminePRB2021,NakamineJPSJ2021}.

From previous NMR results\cite{NakamineJPSJ2019,NakaminePRB2021,NakamineJPSJ2021}, the possible SC states are narrowed down to the $A_{\rm u}$ or $B_{\rm 3u}$ state.
An important method that can be employed to determine which of the two is realized at low $H$ is to measure the Knight shift along the $a$ axis ($K_a$).
However, we could not observe the $a$-axis NMR signal below 20 K in the previous measurement\cite{TokunagaJPSJ2019}.
This is because the nuclear spin-spin relaxation rate $1/T_2$ diverges at approximately 20 K, and the NMR spectrum broadens, which makes observation of NMR signal difficult at low temperatures\cite{TokunagaJPSJ2019}.
To overcome this difficulty, we prepared a $^{125}$Te-enriched single-crystal sample to enhance the NMR signal intensity and succeeded in observing the $a$-axis NMR signal at low temperatures.
We measured $K_a$ down to 75~mK with a field as small as possible ($\mu_0 H \sim 0.56$~T) to avoid $\mbox{\boldmath$d$}$ vector rotation.
We found that, upon subtracting the SC diamagnetic shielding effect, $K_a$ almost follows the normal-state temperature dependence.  
This indicates that the spin susceptibility in the SC state is unchanged as that in the normal state, as expected for a spin-triplet SC state with $\mbox{\boldmath$H$} \perp \mbox{\boldmath$d$}$. 

A single-crystal UTe$_2$ was grown using chemical transport method with iodine as the transport agent\cite{AokiJPSJ2019}. 
Natural U and 99 \% $^{125}$Te-enriched metals were used as the starting materials for the present sample. 
The $^{125}$Te (nuclear spin $I$ = 1/2, gyromagnetic ratio $^{125}\gamma /2\pi$ = 13.454 MHz/T)-NMR measurements were performed on a single crystal of size $1.5 \times 1.6 \times 2.6$ mm$^3$.
We reported that two $^{125}$Te-NMR signals were observed in UTe$_2$ because of the presence of the two inequivalent crystallographic Te sites, as shown in Fig.~\ref{f1}(b)\cite{TokunagaJPSJ2019}. 
Although we measured both signals, there were no qualitative differences in the NMR results.
Thus, we measured the Knight shift in the $^{125}$Te-NMR signal with a larger Knight shift for ensuring the accuracy of the data.  
The frequency-swept NMR spectra were obtained using the Fourier transform (FT) of a spin-echo signal observed after the spin-echo radiofrequency (RF) pulse sequence with a 5 kHz step in a fixed magnetic field. 
The magnetic field was calibrated using a $^{65}$Cu ($^{65}\gamma /2\pi = 12.089$ MHz/T) NMR signal from the NMR coil. 
The NMR spectra in the SC state were recorded using a field-cooling process.
To align the single-crystal sample accurately, the NMR coil with the crystal was mounted on a piezoelectric rotator (ANRv51/ULT/RES+, attocube) with an angular resolution of $\sim 0.1^{\circ}$. 
The sample was rotated in the $ab$ plane and precisely aligned along the $a$ axis using the $H$-angle dependence of the Knight shift\cite{SM_Fujibayashi1}.
The normal-state $K_a$ obtained in the present measurement is consistent with that reported in a recent study\cite{TokunagaJPSJ2022}.
For reliable low-temperature NMR measurements down to 75~mK, the crystal with the NMR coil was immersed in a $^3$He/$^4$He mixture, and the energy of the RF pulses was reduced to ensure that the NMR results were unchanged by the power of the RF pulses. 
Furthermore, we experimentally confirmed the superconductivity just after the NMR RF pulses using a technique reported in a previous study\cite{IshidaJPSJ2020,NakamineJPSJ2019}\cite{SM_Fujibayashi2}. 

\begin{figure}[tbp]
\begin{center}
\includegraphics[width=8cm]{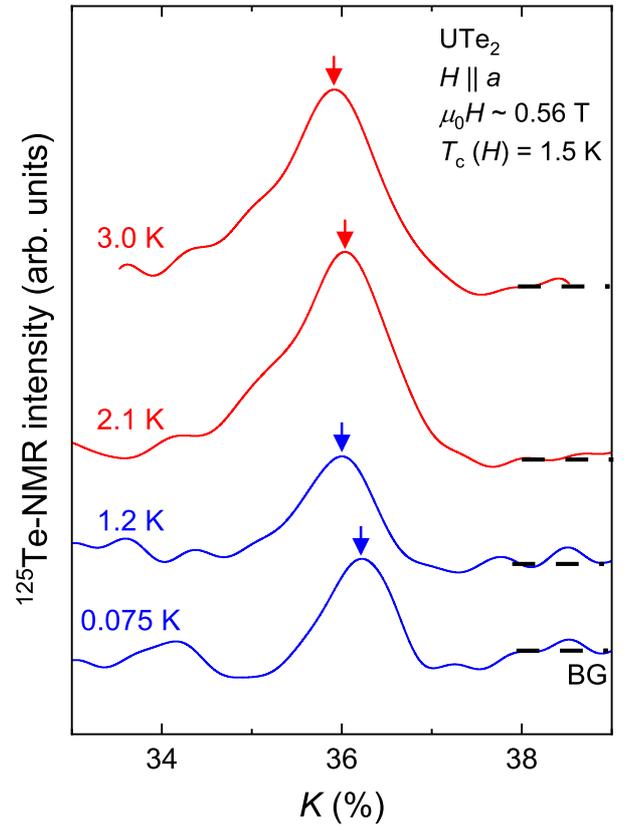}
\end{center}
\caption{(Color online) $^{125}$Te-NMR spectrum measured at various temperatures in $\mu_0 H \sim 0.56$~T when $H \parallel a$. The arrows represent the peaks of the respective NMR spectrum, where the Knight shift was determined. The background level in each spectrum is denoted by BG.
The red (blue) spectrum was measured in the normal (SC) state.}
\label{f2}
\end{figure}
Figure \ref{f2} shows the $^{125}$Te-NMR spectra at several temperatures in $\mu_0H \sim 0.56$ T when $H \parallel a$.
The NMR spectra are shown against $K = (f - f_0 )/ f_0$, where $f$ is the NMR frequency and $f_0$ is the reference frequency; $f_0 = (\gamma_n/2\pi)\mu_0 H$. 
Upon cooling, the NMR spectrum gradually shifts toward a higher $K$.
This is consistent with the result obtained by the heat-up test shown in the supplemental materials\cite{SM_Fujibayashi2}.
In general, the NMR spectrum slightly broadens in the SC state with the SC diamagnetic effect\cite{BrandtPRB1988}, but such a broadening was not observed.
This is due to the anomalous broadening of the NMR line-width below 20 K, as reported recently\cite{TokunagaJPSJ2022}. 
The small SC diamagnetic effect was evaluated with the penetration depth for $H \parallel a$\cite{PaulsenPRB2021}. 

\begin{figure}[tbp]
\begin{center}
\includegraphics[width=8cm]{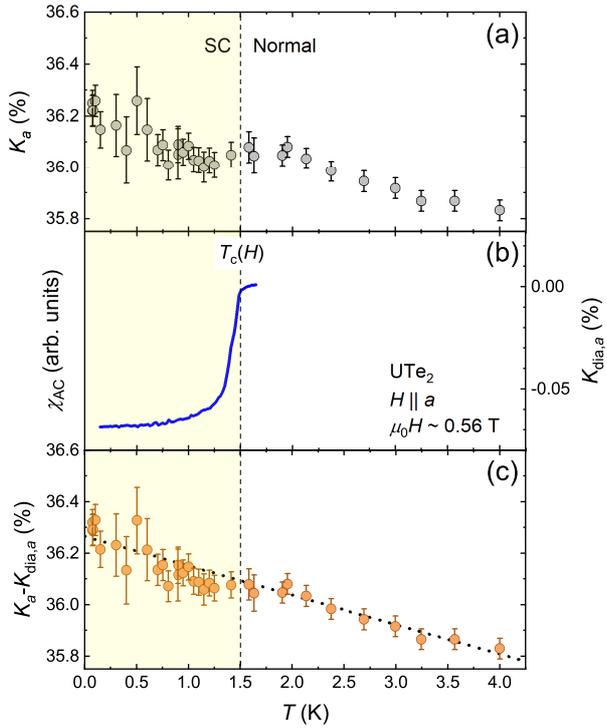}
\end{center}
\caption{(Color online) (a) $T$ dependence of $K_a$ below 4~K. (b) The $T$ dependence of AC  susceptibility measured by the NMR coil at 0.56 T. The right axis indicates $K_{{\rm dia}, a}$ evaluated using Eq.~\ref{eq1}.  (c) $T$-dependence of $K_{{\rm dia}, a}$-term corrected $K_a$ (= $K_a-K_{{\rm dia}, a}$). 
The parameters for the estimation of $K_{\rm dia}$ are listed in Table \ref{t2}.
The dotted line is the linear fit ($A T + B$) of the normal-state Knight shift $K_{\rm N}$ below 4~K.}
\label{f3}
\end{figure}

Figure \ref{f3}(a) shows the $T$ dependence of $K_a$ determined by the peak position of the NMR spectrum in Fig.~\ref{f2}.
Figure \ref{f3}(b) shows the $T$ dependence of AC susceptibility measured in the same field.
Upon cooling, $K_a$ gradually increases in the normal state and slightly decreases below $T_{\rm c}$. 
However, $K_a$ gradually increases again below 1~K while maintaining a similar slope as that obtained in the normal state.

In the SC state, the Knight shift decreases owing to the SC diamagnetic shielding effect, and the value of $K_{{\rm dia}, i}$ at the lowest temperature is approximately expressed as\cite{deGennes} 
\begin{equation} \label{eq1}
K_{{\rm dia}, i} = - \frac{H_{{\rm c1}, i}}{H} \frac{\ln(\frac{\beta \lambda_d}{\sqrt{2.7} \xi})}{\ln{\kappa}}.
\end{equation}
Here, $\xi$ is the Ginzburg-Landau (GL) coherence length; $\beta$ is a factor that depends on the vortex structure and is 0.38 for the triangular vortex lattice; $\lambda_d$ is the distance between the vortices and is calculated using the relation $\phi_0 = \frac{\sqrt{3}}{2} \lambda_d^2(\mu_0H_{\rm ext})$; and $\kappa$ is the GL parameter.
We estimated $K_{{\rm dia}, a}$ with the SC parameters reported by Paulsen {\it et al.}\cite{PaulsenPRB2021}, which are listed in Table \ref{t2}, and the $T$ dependence of $K_{{\rm dia}, a}$ was estimated from the experimental results of the AC susceptibility shown in Fig.~\ref{f3}(b).
\begin{table}[htb]
\begin{center}
\caption{\label{t2}Superconducting parameters used for the estimation of $K_{\rm dia}$ at low temperatures. The parameters were obtained from the reference\cite{PaulsenPRB2021}. $H_{\rm ext}$ is the external field used for the Knight-shift measurement. }  
\vspace{5mm}
  \begin{tabular}{ccccccc}\hline \hline 
      & & $H \parallel a$ & & $H \parallel b$ & & $H \parallel c$ \\ \hline
$\mu_0 H_{\rm ext}$ (T) & & 0.56 & & 1 & & 1 \\      
$\mu_0 H_{\rm c1}$ (mT) & & 1.5  & & 2.5 & & 1.3 \\
$\kappa$ & & 200 & & 55 & &45\\
$\xi$ ($H \parallel i$) (nm) & & 3.85 & & 7.1 & & 7.8 \\
$K_{\rm dia} (\%)$ & & 0.068 & & 0.028 & &0.012 \\ \hline \hline
  \end{tabular}
  \end{center}
\end{table}
Using the evaluated $K_{{\rm dia}, a}$, the $T$ dependence of $K_a$, in which $K_{{\rm dia}, a}$($T$) is corrected, is shown in Fig.~\ref{f3}(c). 
It was found that the small decrease below $T_{\rm c}$ can be explained by the SC diamagnetic effect; after subtracting the contributions from the SC diamagnetic effect, the anomaly at $T_{\rm c}$ becomes negligibly small, as shown in Fig.~\ref{f3} (c). 
To investigate the $T$ dependence of spin susceptibility, we fit $K_a$($T$) in the normal state below 4.2 K to a linear function and subtracted this temperature dependence from the experimental data.  
\begin{figure*}[t]
\begin{center}
\includegraphics{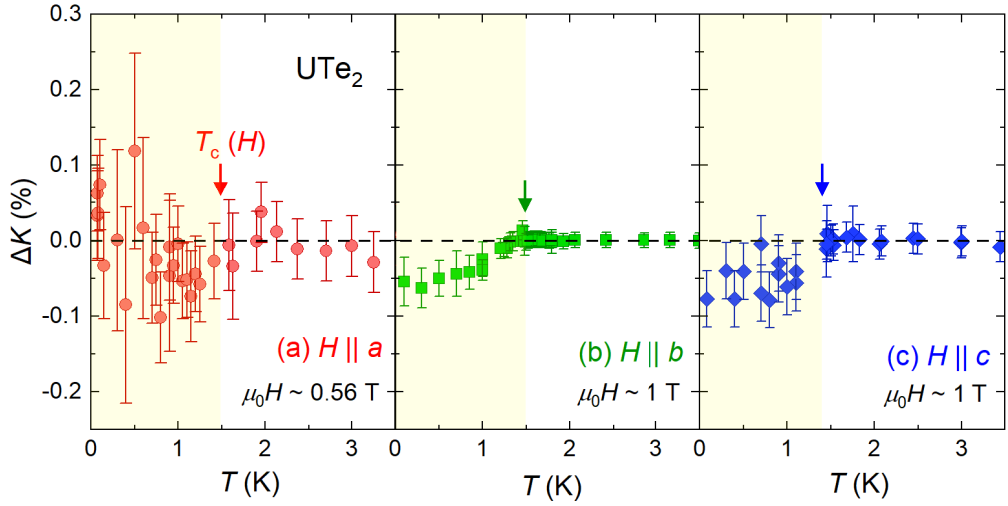}
\end{center}
\caption{(Color online) $T$ dependence of $\Delta K_a$, in which the normal-state $T$ dependence of $K_{\rm N}$($T$) were subtracted from $K_a-K_{\rm dia}$(see in text). 
The $T$ dependence of $K_b$ and $K_c$ are quoted from the references\cite{NakaminePRB2021,NakamineJPSJ2021,NakamineDcTh}. 
The temperature dependence of the normal-state Knight shift $K_{\rm N}$ in both axes was subtracted from $K_i-K_{{\rm dia}, i}$ ($i$ = $b$ and $c$)\cite{SM_Fujibayashi3}.  }
\label{f4}
\end{figure*}

Figure \ref{f4} (a) shows the $T$ dependence of the corrected $\Delta K_a$, in which the normal-state $T$ dependence $K_{{\rm N}, a}$($T$) is subtracted from $K_a-K_{{\rm dia}, a}$ ($\Delta K_a \equiv K_a - K_{{\rm dia}, a} - K_{{\rm N}, a}$).
$\Delta K_a$ corresponds to the variation of the spin component of the Knight shift ascribed to the SC transition, and $\Delta K_i$ ($i$ = $b$ and $c$) is estimated in the same manner, as shown in Figs.~\ref{f4}(b) and \ref{f4}(c) for comparison\cite{NakaminePRB2021,NakamineJPSJ2021}. 
Note that the spin part of the Knight shift in all directions ascribed to superconductivity was roughly estimated as $\sim 0.6$ \%  from the change in the electronic term of the specific heat in the SC state with assuming a free-electron spin\cite{aoki2021condmat}. 
Although the experimental data in the SC state are scattered owing to the reduction in the intensity of the NMR spectrum, it was found that $\Delta K_a$ does not decrease at the lowest temperature and is almost unchanged in the SC state.
This behavior contrasts with the behavior of $\Delta K_b$ and $\Delta K_c$, in which an appreciable decrease was observed in the SC state.
Furthermore, if we consider the large anisotropy of the normal-state spin susceptibility, we conclude from the present results that the spin susceptibility when $H \parallel a$ is almost unchanged in the SC state.   

From the theoretical study, the contrasting behavior of $\Delta K_a$ in the SC state between the $A_{\rm u}$ and $B_{\rm 3u}$ states was pointed out by Hiranuma and Fujimoto\cite{HiranumaJPSJ2021}.
They calculated the spin susceptibility in the SC state based on the multiband orbital $f$-electron bands of total angular moments $j$ = 5/2 with spin-orbit couplings for an orthorhombic structure. 
They demonstrated that for $H$ parallel to the magnetic hard axis, the spin susceptibility ($\chi_{bb}$) decreases to 90 \% of the normal-state spin susceptibility $\chi_{\rm N, spin}$ at 0~K and the Pauli depairing effect is suppressed in the $A_{\rm u}$ and $B_{\rm 3u}$ states.
The decrease amounts of the spin susceptibility are quantitatively in good agreement with the experimental observations.\cite{NakamineJPSJ2019,NakaminePRB2021,NakamineJPSJ2021,aoki2021condmat}
Additionally, they showed that the spin susceptibility parallel to the magnetic easy axis ($\chi_{aa}$) significantly decreases in the unitary $A_{\rm u}$ and non-unitary $A_{\rm u} + B_{\rm 3u}$ states, but does not decrease in unitary $B_{\rm 3u}$ state for $\mbox{\boldmath$H$} \perp \mbox{\boldmath$d$}$, and that the decrease in $\chi_{aa}$ in the $A_{\rm u}$ and $A_{\rm u} + B_{\rm 3u}$ states is approximately three times larger than that in $\chi_{bb}$.
This is because $\chi_{aa}$ consists of the intra-band contribution, and unlike in the case of $\chi_{bb}$, there are no inter-band terms that do not decrease below $T_{\rm c}$.
Therefore, present result strongly suggests that the predominant $\mbox{\boldmath$d$}$ vector is a unitary $B_{\rm 3u}$ state.
The realization of the $B_{\rm 3u}$ state in the $H$ = 0 or low-$H$ SC region was also indicated by the microscopic DFT + U  calculations\cite{IshizukaPRB2021, ShishidouPRB2021}. 
Note that there remains a possibility of a $\mbox{\boldmath$d$}$-vector rotation from $A_{\rm u}$ at zero field to $B_{\rm 3u}$ at the field smaller than 0.56~T.
However, because the measurement field for $H \parallel a$ is much lower than the $\mbox{\boldmath$d$}$-vector rotation field in $H \parallel b$ (12.5~T) and $c$ axis (5.5~T)\cite{NakaminePRB2021,NakamineJPSJ2021}, this possibility is unlikely.

Finally, we comment on the possibility of the non-unitary state.
Our results cannot exclude the possibility of a non-unitary state, including the tiny $A_{\rm u}$ or $B_{\rm 2u}$ state, because the expected decrease in $K_a$ would be small in such non-unitary states.
However, it seems that $\Delta K_a$ slightly increases below 1~K, although the data points are largely scattered and inconclusive. 
It is considered that the spin susceptibility might slightly increase due to the non-unitary spin-triplet pairing, in which the two-spin population becomes unequal at low temperatures.
This possibility has already been pointed out theoretically\cite{AndersonPR1961,MiyakeJPSJ2014}.
To confirm this possibility, further experiments, particularly in higher $H$ are desired.
Note that the 2nd-order transition from the paramagnetic state to the non-unitary state at zero field is theoretically forbidden from the symmetry arguments\cite{Mineev2008JPSJ2008}.
Thus, the double transition is inevitably required at zero field, however, recent high-quality crystals of UTe$_2$ seem to exclude the possible double transition\cite{RosaCondMat2021, aoki2021condmat}.

In conclusion, we investigated the spin susceptibility along the $a$ axis in the SC state of UTe$_2$ by $^{125}$Te-NMR measurements.
Based on the present and previous Knight-shift results\cite{NakamineJPSJ2019,NakaminePRB2021,NakamineJPSJ2021} and theoretical calculations \cite{HiranumaJPSJ2021}, the dominant SC pairing symmetry is determined to be the $B_{\rm 3u}$ state, in which the main spin component of the SC pairing is along the $a$ axis. 
This SC state is consistent with the spin anisotropy in the normal state.
The important issue is how the SC spin degrees of freedom behave under various experimental circumstances, which is now investigated. 

The authors would like to thank J. Ishizuka, Y. Yanase, K. Machida, S. Fujimoto, V. P. Mineev, Y. Maeno, S. Yonezawa, J-P. Brison, G. Knebel, and J. Flouquet for their valuable inputs in our discussions. 
This work was supported by the Kyoto University LTM Center, Grants-in-Aid for Scientific Research (Grant No. JP15H05745, JP17K14339, JP19K03726, JP16KK0106, JP19K14657, JP19H04696, JP19H00646, and JP20H00130).




\vspace{1cm}
\noindent
Supplemental Materials of \\
{\bf ``Superconducting order parameter in UTe$_2$ determined by Knight-shift measurement''}

\section{Sample alignment}
\begin{figure}[h]
\begin{center}
\includegraphics[width=8cm]{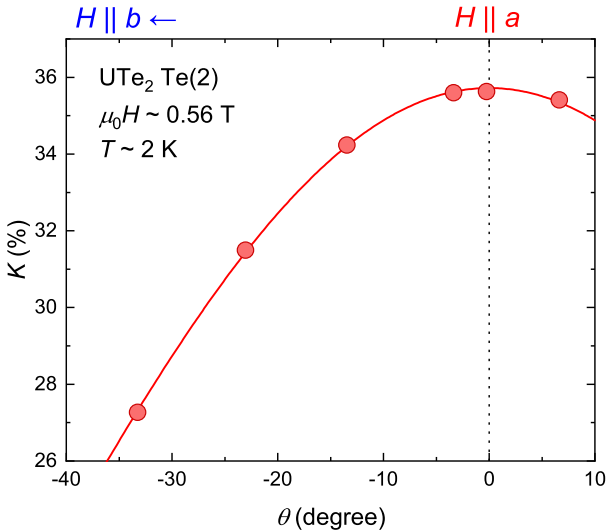}
\end{center}
\caption{(Color online) Angle dependence of the Knight shift of the $^{125}$Te - NMR signal with the larger Knight shift when the single-crystal sample was rotated in the $ab$ plane. $\theta = 0$ is determined by fitting the experimental data to the theoretical function of $K(\theta) = K_a \cos^2\theta + K_b \sin^2\theta$. }
\label{fs1}
\end{figure}

To apply the magnetic field parallel to the $a$ axis, we rotated the single-crystal sample in the $ab$ plane under $\mu_0 H \sim 0.56$ T,  and measured the angle dependence of the Knight shift as shown in Fig.~\ref{fs1}.
We determined the $a$ axis as the angle at which the Knight shift is maximum.

\begin{figure}[h]
\vspace{5mm}
\begin{center}
\includegraphics[width=8cm]{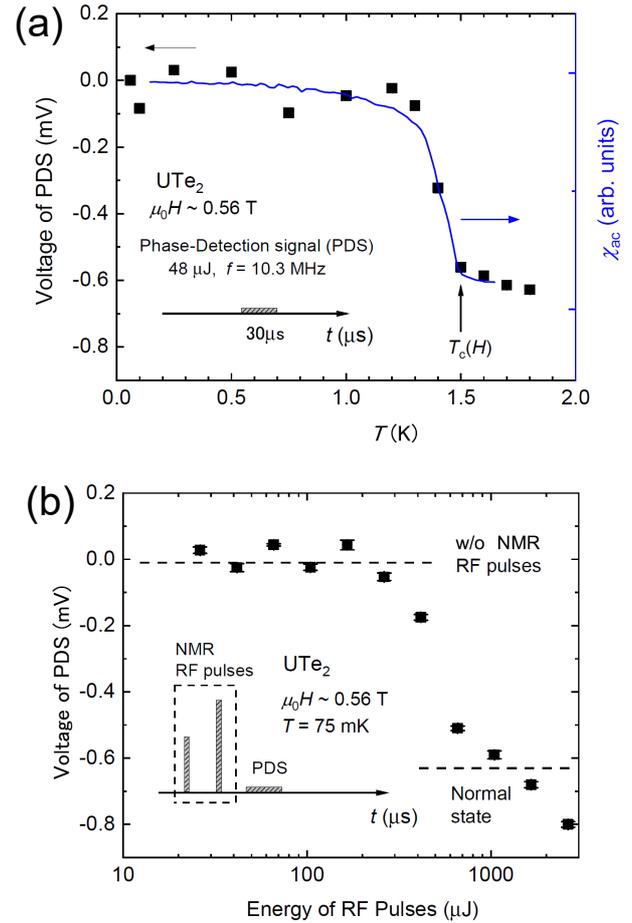}
\end{center}
\caption{(Color online) (a) Temperature dependence of the voltage of the imaginary part of the output signal from the NMR receiver after a phase-detection signal (PDS) was inputted. The pulse width and the output power of the PDS are 30 $\mu$s and 1.6~W, respectively.
(b) the dependence of the voltage of the PDS on the total energy of two RF pulses applied before the PDS in the same condition as the $K_a$ measurement.
The dependeence was measured in the $^3$He/$^4$He mixture temperature of 75~mK. The pulse sequence of the test is shown in the inset of Fig. \ref{fs2}(b). 
The voltage shown by ``w/o NMR RF pulses" is the voltage of the PDS without NMR RF pulses, and the voltage shown by ``Normal state'' is the voltage of the PDS expected in the normal state.} 
\label{fs2}
\end{figure}
\section{Heating Effect by NMR RF-pulses}
The superconductivity just after the RF pulses for the NMR-signal observation was confirmed by the following heat-up test using the same set-up as the NMR measurements. 
We applied a weak RF pulse with the pulse width of 30 $\mu$s and the same frequency of 10.3 MHz as the NMR measurements, and measured the voltage of the imaginary part of the output signal from the NMR receiver.
We call this weak RF pulse ``phase-detection signal (PDS)'', and the RF-pulse energy was 48 $\mu$J.
Here, the energy was just expressed by the product of nominal output value of the NMR power amplifier and the pulse width, as the estimation of the RF-pulse energy absorbed directly by the sample is difficult.

First, we measured the $T$ dependence of the voltage of the imaginary part of the output signal of the PDS. 
In the measurement, we applied $\mu_0 H \sim 0.56$ T in $H \parallel a$ as in the present NMR measurement.   
Fig.~\ref{fs2}(a) shows the $T$ dependence of the voltage.
The voltage changed at $T_{\rm c}$ owing to the change of the impedance of the NMR tank circuit since the inductance of the NMR coil with the sample was changed in the SC state.
As shown in Fig.~\ref{fs2}(a), the $T$ dependence of the voltage is similar to that of $\chi_{\rm AC}$ measured by the NMR coil, indicating that the voltage of PDS is related to $\chi_{\rm AC}$.

Next, we applied two RF pulses corresponding to $\pi/2$ and $\pi$ pulses just before the PDS. 
The pulse width of each RF pulse is fixed 7 $\mu$s and the voltage of the $\pi$/2 pulse is half of that of $\pi$ pulse. 
The pulse sequence of the test is shown in the inset of Fig.~\ref{fs2}(b).
We measured the dependence of the voltage of the PDS on the total energy of two RF pulses in the same condition as the $K_a$ measurement in the $^3$He/$^4$He mixture temperature of 75~mK.  
This measurement corresponds to the instantaneous measurement of $\chi_{\rm AC}$ at the spin echo position just after the NMR RF pulses with the various energies. 
Figure \ref{fs2}(b) shows that the voltage of the PDS is almost unchanged up to 300 $\mu$J and that the superconductivity was destroyed by the RF pulses with $\sim$ 1000 $\mu$J.  

\begin{figure}[h]
\begin{center}
\includegraphics[width=8cm]{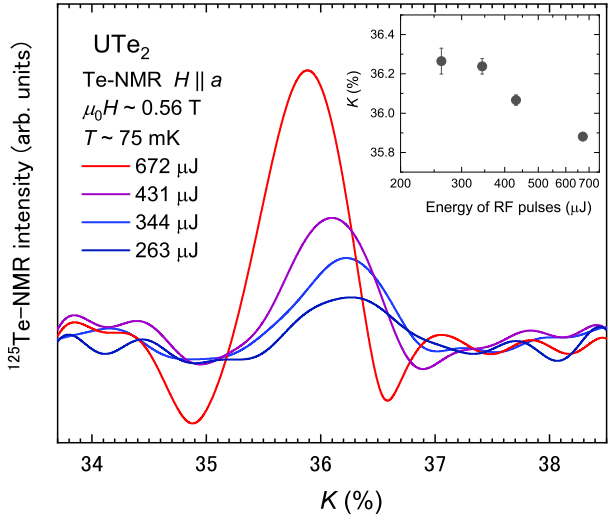}
\end{center}
\caption{(Color online) The $^{125}$Te-NMR spectrum measured with different RF-pulse energy. 
The RF-pulse energy was expressed by the product of nominal output value of the NMR power amplifier and the pulse width.
The Te-NMR spectrum by the smaller energy than 263 $\mu$J was comparable to the noise level.
The inset is the RF-pulse energy dependence of the Knight shift determined from the peak frequency of the $^{125}$Te-NMR spectrum.
The Knight shift was almost unchanged when the NMR spectrum was measured by the RF pulses with the energy smaller 344 $\mu$J.}
\label{fs3}
\end{figure}
We also checked the heat-up effect in the NMR measurement.
We investigated the dependence of $^{125}$Te-NMR spectrum on the energy of the applied RF pulses at low temperatures in 0.56 T.
In the measurement, the $^3$He/$^4$He mixture was fixed at 75 mK, and the Te-NMR spectrum was measured by the RF pulses with different energies, which is shown in Fig.~\ref{fs3}.
With increasing the RF-pulse energy, the NMR spectrum shifts towards a lower-$K$.
This is due to an instantaneous heat-up effect of the sample and is similar to the temperature dependence.    
It was found that the peak frequency of the NMR spectrum was almost unchanged when the NMR spectrum was measured by the RF pulses with the energy smaller than 344 $\mu$J.
Thus the NMR spectrum at various temperatures in the paper was measured with the RF-pulses with 344 $\mu$J.

\begin{figure}[h]
\begin{center}
\vspace{5mm}
\includegraphics[width=7.5cm]{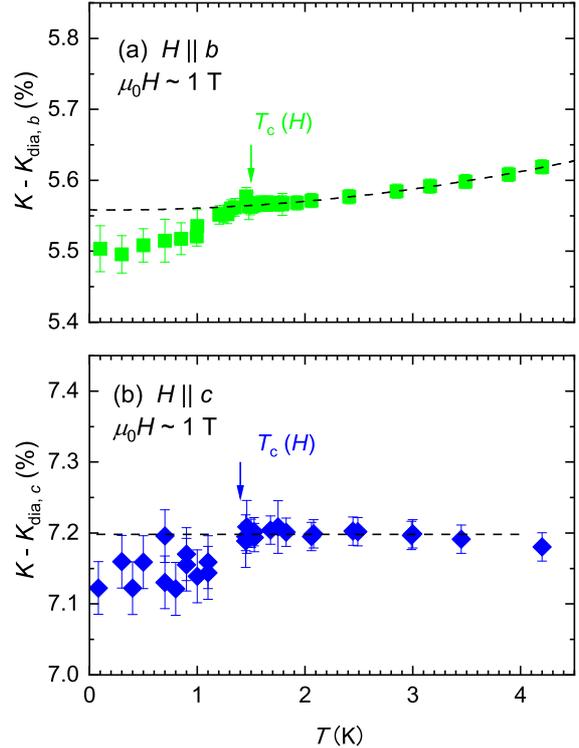}
\end{center}
\caption{(Color online) Temperature dependence of $K_b$ and $K_c$ below 4.5~K. The SC diamagnetic effect in Knight shift $K_{\rm {dia}}$ below $T_{\rm c}$ was corrected in $K_b$ and $K_c$. The dotted curve in (a) is the quadratic fitting function of the normal-state Knight shift below 4.5 K, and the dotted line in (b) is a constant value ($K_c$ =7.2 \%), as the normal-state $K_c$ is almost $T$ independent below 3.5 K.
These dotted functions ($K_{\rm N}$) are subtracted from $K - K_{{\rm dia}, i}$ ($i$ = $b$ and $c$) in Fig.~\ref{f4}(b) and (c) in the text. }
\label{fs4}
\end{figure}
\section{Temperature dependence in the normal-state Knight shift}
Figure \ref{fs4} shows the temperature dependence of $K_b$ and $K_c$. 
In the SC state, the SC diamagnetic effect $K_{\rm dia}$ is corrected: the temperature dependence of $K_{\rm dia}$ was estimated from the AC susceptibility measurement in the SC state and the magnitude of $K_{\rm dia}$ at low temperature was evaluated with the physical parameters shown in Table II in the main paper. 
The temperature dependence of the normal-state $K_b$ was fitted to the quadratic function, and this component was subtracted from $K_b - K_{{\rm dia},b}$ in Fig.~\ref{f4}(b) in the paper.
As for $K_c$, the constant value (= 7.2 \%) was just subtracted in Fig.~\ref{f4}(c) in the paper, as the normal-state $K_c$ is almost $T$ independent below 3.5 K.

\end{document}